\documentclass[pageno]{jpaper}

\usepackage[ruled,linesnumbered]{algorithm2e}
\usepackage[normalem]{ulem}
\usepackage[inkscapelatex=false]{svg}
\usepackage{amsmath}
\usepackage{threeparttable}
\begin{document}

\author{Peng Liang, Hao Zheng, \textsuperscript{*}Teng Su, Linbo Qiao, Dongsheng Li
\\ \normalsize National University of Defense Technology, \textsuperscript{*}Huawei Technologies Co. Ltd
\\ \{peng\_leung, zhengh, qiao.linbo, dsli\}@nudt.edu.cn
\\ \textsuperscript{*}suteng@huawei.com
\thanks{A preprint version, change at any time.}
}
\title{
TAPS: Topology-Aware Intra-Operator Parallelism Strategy Searching Algorithm for Deep Neural Networks}

\date{}
\maketitle

\thispagestyle{empty}

\begin{abstract}
TAPS is a Topology-Aware intra-operator Parallelism strategy Searching algorithm that generates intra-operator parallelism strategies by considering both intra-node and inter-node bandwidth.
Most of the existing auto-parallelism works use the communication volume as the communication cost directly when generating strategies, which we prove to be sub-optimal in multi-nodes cases.
We design a topology-aware cost model for multi-node intra-operator parallelism strategy searching. Numerical experiments demonstrate that TAPS can generate strategies with up to 85\% fewer communication costs, which outperform the latest baselines.
\end{abstract}

\section{Introduction}
Large-scale Deep Learning (DL) models have been a huge hot topic in recent years for their great performance improvements in fields like \cite{brown2020language, deeplearning, alphafold}, which is a result of scaling up model sizes and dataset sizes. 
For example, PaLM with 540 billion parameters is trained with a corpus of 780 billion tokens that represent a wide range of natural language use cases \cite{PaLM}.

As the model size significantly increases, training models with a single device or even within a node is no longer practical.
Thus, researchers use distributed deep learning to train these models \cite{distributed2020survey}.
Manual strategies like \cite{megatron2019} have been widely used in training transformer-based models for their good performance.
However, it is often not optimal because optimal parallelism strategies vary when the model or training environment changes, in which case researchers and engineers may need to redesign strategies.

To relieve us from the parallelism design procedure, researchers propose auto-parallelism algorithms \cite{cai2021tensoropt, optcnn, zheng2022alpa} that can find decent strategies given a specific model and environment.
These algorithms first model parallelism strategies' communication costs and then use a dynamic programming or an integer linear programming (ILP) method to find the optimal strategy. 

As model size grows larger, a single node can no longer hold an entire large-scale model.
Thus, using multi-nodes to train a model becomes necessary.
Our key observation is that in a multi-node environment, the bandwidth within a node (intra-node bandwidth) and across nodes (inter-node bandwidth) are different, and the intra-node bandwidth is much higher than inter-node bandwidth in most cases.
However, existing searching algorithms model the communication cost using the communication volume directly, ignoring the difference between the bandwidths and resulting in sub-optimal strategies.
Based on this observation, we propose a topology-aware parallelism strategy searching algorithm called TAPS, which can capture the difference between intra-node and inter-node communication and thus generates better parallelism strategies.

We first construct a topology-aware cost model, which can determine the inter-node communication times as well as the topology-aware communication cost given a communication axis of a tensor.
Then we formalize the strategy searching problem as an integer linear programming problem, after which we use a third-party solver to solve the final strategy decision.

In summary, we make the following contributions:
\begin{itemize}
  \item We prove that the volume-based communication cost model is insufficient to generate optimal intra-operator parallelism strategy in multi-nodes cases.
  \item We provide a heuristic solution in optimizing tensor redistribution sequences.
  \item We analyze the communication in multi-node environments and propose a topology-aware communication cost model, which can calculate more accurate communication costs of a parallelism strategy of an operator.
  \item We design and implement TAPS, a strategy-searching algorithm that works for distributed DL.
  \item We numerically evaluate TAPS on several models of different configurations. We compare TAPS with volume-based searching. Our experiments show that TAPS can find strategies with up to 85\% fewer communication costs.
\end{itemize}

\section{Background}
\subsection{Existing Parallelism Methods}
Since Hinton \cite{alexnet} trained AlexNet using two GPUs in 2012, researchers have proposed many parallelism methods, including data parallelism (DP), model parallelism (MP), and pipeline parallelism(PP). 
\subsubsection{Data Parallelism}
Data parallelism partition and distribute the data across devices that has a replicated model.
Each device computes the gradients using the split data and uses communication like AllReduce or Broadcast to synchronize the gradients or model parameters with other devices. So that after every iteration, the models on all workers are the same.

\subsubsection{Model Parallelism}
Model parallelism partition the model parameters across devices and make devices process the same data. Model parallelism produces partial-sum or sliced results when the parameter matrix is partitioned row-wisely and column-wisely, respectively. Row-wise MP (Row-MP) requires synchronization to unify the operator's results on different devices. Column-wise MP (Column-MP) does synchronization only in backward propagation. 

\subsubsection{Pipeline Parallelism}
Pipeline parallelism partition operators in a model into several stages and let devices hold only one or a few of them. Meanwhile, PP splits a mini-batch of data into several micro-batches and feeds them one by one into the first stage. When a stage finishes its computation, it sends the result to its next stage. Different stages can be handled simultaneously; thus, PP forms a pipeline that can improve performance.

\subsection{Intra- and Inter-Operator Paralleism}
Alpa \cite{zheng2022alpa} catalog existing parallelism methods into two orthogonal categories: intra-operator and inter-operator parallelism. 
Intra-operator parallelisms are parallelism schemes that partition an operator's involved tensors along some dimensions, assign the resulting partitioned computation to multiple devices, and let them execute different parts of the computation simultaneously. From this view, we can treat data parallelism as a scheme that partitions an operator's input and output tensor along the batch-size axis; we can treat Row-MP as a scheme that partitions an operator's input and weight tensor along the channel-in axis; we can treat Column-MP as a scheme that partitions weight tensor and output tensor along the channel-out axis. Inter-operator parallelism, including pipeline parallelism, partitions models into several stages with multiple operators.

This paper focuses on generating multi-dimensional intra-operator parallelism strategies in multi-node environments. 

\subsection{Strategy Searching Algorithm}
Researchers have proposed methods to search parallelism strategies automatically. ToFu\cite{wang2019tofu}, TensorOpt\cite{cai2021tensoropt}, and Alpa\cite{zheng2022alpa} generate intra-operator parallelism strategies by minimizing the overall communication cost of a computation graph under the observation that all different strategies of an operator have the same computation cost. ToFu and TensorOpt adapt the dynamic programming algorithm that OptCNN\cite{optcnn} propose to produce better results. Alpa formalizes the searching problem as an integer programming problem and uses a solver to handle the solution progress. However, they assume the bandwidths of clusters are equal everywhere, ignoring the difference between the intra-node bandwidth and inter-node bandwidth. This assumption may limit the searching algorithm to find the optimal strategies, as, in large-scale clusters,  intra-node bandwidth is much higher than inter-node bandwidth. In this paper, we propose a topology-aware communication cost model aware of the intra-node and inter-node bandwidth, which helps generate more fine-grained strategies.

\section{Overview}
TAPS is an algorithm that generates intra-operator parallelism strategies by minimizing the communication cost of the computation graph. TAPS takes a computation graph $G=(V, E)$ and device graph $D=(V_D,  E_D)$ as inputs, and output a partition set $P$, which consists of strategy decisions of every operator $v_i \in V$ in $G$. The computation graph contains operator information, like shapes and operator types. The device graph indicates the device types and the bandwidth between devices. 
TAPS gives a solution in two steps: First, TAPS creates an auxiliary graph where each node indicates an operator with a specific strategy and computes the weights for each edge $(u, v)$ in the auxiliary graph, which equals the intra-operator communication cost of $v$ plus tensor redistribution communication cost between $u$ and $v$. Then, TAPS formalizes the searching problem as an integer linear programming problem using the information in the auxiliary graph and uses a third-party solver to solve the optimal strategy.

\section{Communication Cost Model}
In this section, we give the details of our topology-aware communication cost model. We first illustrate the details of the volume-based cost model. Based on the volume-based cost model, we calculate the corresponding topology-aware communication cost using the volumes and effective bandwidth.

\subsection{Volume-based cost model}
Previous works \cite{cai2021tensoropt, Accpar, DoubleRec} model the communication cost of each strategy by symbolically computing the their communication volume. The communication volume of an operator consists of intra-operator communication and inter-operator communication. Intra-operator communication reduces the partial sums generated in computing. Inter-operator communication transforms tensor to fit the succeeding operator's strategy.

\subsubsection{Intra-operator communication}
Taking MatMul as an example, its forward computation is shown as Eq.\ref{eq: forwardcomputation}, and its backward computation is shown as Eq.\ref{eq: deltaweight} and Eq.\ref{eq: error_x}. 
\begin{align}
  Y &= X W \label{eq: forwardcomputation} \\ 
  \delta W &= X^T E_y \label{eq: deltaweight} \\
  E_x &= E_y W^T  \label{eq: error_x}
\end{align}
Let $d$, $r$, $c$ denote the data parallelism (DP)\cite{PyTorch-DDP}, Row-MP, and Column-MP \cite{megatron2019} degrees of a MatMul operator, respectively; $p=drc$ denotes the total device number and is the power of 2. Then we split the $X$ and $W$ matrices like:
$${\begin{matrix}
X= \begin{bmatrix}
    X_{11} &X_{12} & \dots &X_{1r} \\
    X_{21} &X_{22} & \dots & \vdots \\
    \vdots &\vdots & \ddots & \vdots \\
    X_{d1} &X_{d2} & \dots & X_{dr}
    \end{bmatrix} ,& 
    W= \begin{bmatrix}
      W_{11} &W_{12} & \dots &W_{1c} \\
      W_{21} &W_{22} & \dots & \vdots \\
      \vdots &\vdots & \ddots & \vdots \\
      W_{r1} &W_{r2} & \dots & W_{rc}
      \end{bmatrix}.
    \end{matrix}}   
$$
\begin{figure*}
  \centering
  \includesvg[width=\linewidth]{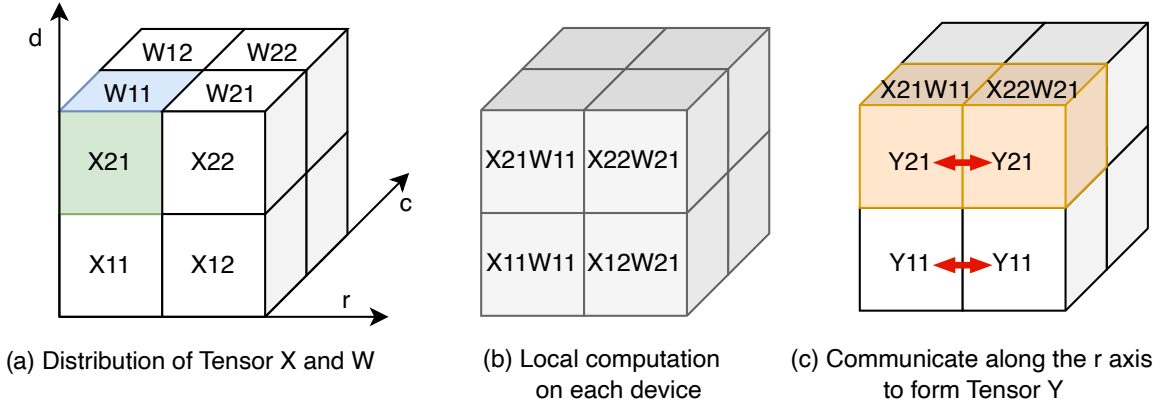}
  \caption{Multi dimensional Intra-P of a MatMul operator on 8 devices where $d=r=c=2$}
  \label{fig: 3d-tp}
\end{figure*}
After splitting the matrices $X$ and $W$, we distribute their sub-blocks to corresponding devices. As Figure \ref{fig: 3d-tp}.(a) shows, where each cube represents a device, each sub-block of $X$ is replicated along axis $c$, and $W$ is replicated along axis $d$. As Figure \ref{fig: 3d-tp}.(b)(c) shows, we then compute the local results of $Y$ on each device and communicate them to form the final matrix $Y$. The communication is a reduction operation of local results and is mathematically equivalent to Eq. \ref{eq: accumulation}.
\begin{equation}
  \label{eq: accumulation}
  Y_{ij} = \sum_{k=1}^{r} X_{ik} W_{kj}
\end{equation}
Final matrix $Y$ is split like: 
$${\begin{matrix}
  Y= \begin{bmatrix}
      Y_{11} &Y_{12} & \dots &Y_{1c} \\
      Y_{21} &Y_{22} & \dots & \vdots \\
      \vdots &\vdots & \ddots & \vdots \\
      Y_{d1} &Y_{d2} & \dots & Y_{dc}
      \end{bmatrix} ,
      
      \end{matrix}}
  $$
where each sub-block $Y_{ij}$ is replicated along $d$ axis.

Suppose we are using a bandwidth optimal Ring-AllReduce algorithm \cite{optimalRingAllreduce}, the communication volume of a MatMul operator accumulating results of $Y$ on each device (i.e., the volume of Row-MP) is: 
\begin{equation}
  V_{AR}^Y = \frac{2(device\_num -1)}{device\_num}data\_size = \frac{2(r-1)b\ out}{drc}.
\end{equation}
Similarly, we give the communication volume of DP and Column-MP in a Matmul operator by computing the communication volume of acuumulating results of $\delta W$ and $E_X$, respectively, which are:
\begin{equation}
  V_{AR}^{\delta W} = \frac{2(d-1)in\ out}{drc},
\end{equation}
\begin{equation}
  V_{AR}^{E_X} = \frac{2(c-1)b\ in}{drc}.
\end{equation}
Finally, the overall communication volume of a MatMul operator is: 
\begin{equation}
   Volume = \frac{2((d-1)in\ out + (r-1) b\ out + (c-1) b\ in)}{drc}
\end{equation}
\subsubsection{Inter-operator communication} 
Inter-operator communication happens when there are tensor redistributions between two operators. Tensor redistributions are sequences that consist of several redistribution operators like All-Gather, Slice, and All-To-All. In this subsection, we give our solution for generating proper redistribution operator sequences.

Let $O_{out}, O_{in}$ denote two operators and $T$ denote the output $n$-dimensional tensor of $O_{out}$ and the input tensor of $O_{in}$. $S_T = [s_0, s_1, ..., s_{n-1}]$ is the shape of $T$ before partition. Suppose the depths of device matrix of $O_{out}$ and $O_{in}$ is $h_{out}$ and $h_{in}$. The device matrix in $O_{out}$ and $O_{in}$ are $\mathcal{D}_{out} = [d_{out,h_{out}-1}, d_{out,h_{out}-2}, ..., d_{out, 0}]$ and $\mathcal{D}_{in} = [d_{in,h_{in}-1}, d_{in, h_{in}-2}, ..., d_{in, 0}]$, respectively. The tensor maps of $T$ in $O_{out}$ and $O_{in}$ are $M_{in} = [m_{out, 0}, m_{out, 1}, m_{out, n-1}]$ and $M_{out} = [m_{in,0}, m_{in,1}, ..., m_{in, n-1}]$, respectively. To do the tensor redistribution, the device matrices and tensor shapes of $O_{in}$ and $O_{out}$ must be the same. We unify them by two steps. In step 1, we unify device matrices by factorizing some dimensions in two device matrices, which may result in a shape inconsistency of $T$ in two operators. Thus in step 2, we need to unify the tensor shape under the unified device matrix additionally. Note that the two-step unification does not change the physical distribution of a Tensor. Table \ref{tab: unifying} shows an example of unifying a 2-dimensional tensor between $O_{out}$  and $O_{in}$. In step 1, we factorize "8" in two device matrices and replace them by the factorizing results $[4, 2]$ and $[2, 4]$ for $O_{out}$ and $O_{in}$, respectively. Meanwhile, we must change the tensor maps and shapes as we modify device matrices. Since the tensor shapes change in step 1, we need to unify it again before we infer tensor redistribution operators. In step 2, we reshape the tensor in $O_{in}$ and $O_{out}$ to make them have the same shape and modify tensor maps simultaneously. 

\begin{table*}[]
  \centering
  \caption[]{Unifying Device Matrix and Tensor Shape of Tensor $T$} \label{tab: unifying}
  \begin{tabular}{c|c|c|c|c}
  \hline
  \textbf{Step} &\textbf{Operator} &\textbf{Device Matrix} &\textbf{Tensor Map} &\textbf{Tensor Shape} \\ \hline
  \multirow{2}{*}{0: Initial} & $O_{out}$       & $[2, 8]$ & $[1, 0]$        & $[s_0, s_1]$      \\ \cline{2-5} 
                           & $O_{in}$        & $[8, 2]$ & $[1, 0]$        & $[s_0, s_1]$      \\ \hline \hline 
  \multirow{2}{*}{1: Unifying device matrix} &
    $O_{out}$ &
    \multirow{2}{*}{$[2, 4, 2]$} &
    $[2, 1, 0]$ &
    $[s_0, 4, s_1/4]$ \\ \cline{2-2} \cline{4-5}
                           & $O_{in}$ &          & $[2, 1, 0]$     & $[2, s_0/2, s_1]$ \\ \hline \hline
  \multirow{2}{*}{2: Unifying tensor shape} &
    $O_{out}$ &
    \multirow{2}{*}{$[2, 4, 2]$} &
    $[2, -1, 1, 0]$ &
    \multirow{2}{*}{$[2, s_0/2, 4, s_1/4]$} \\ \cline{2-2} \cline{4-4}
                           & $O_{in}$  &          & $[2, 1, 0, -1]$ &                   \\ \hline
  \end{tabular}
  \end{table*}

After unifying the device matrix and tensor shape, we can infer the redistribution operators. A naive way to do the redistribution is to AllGather along all the workers and then partition along axes that are not repetitive. To reduce the communication cost, we use a heuristical algorithm \ref{alg:optimized Tensor redistribution} to generate tensor redistribution operators. Our algorithm contains three optimizations. First, we only AllGather along the necessary axes of the tensor, which are partitioned in $O_{out}$ and replicated in $O_{in}$. Second, we rearrange the redistribution sequence, putting dependent Slice before AllGather to reduce the communication volume that AllGather produces. Third, we replace the implicit permutations (i.e., AllGather and Slice along the same axis in the device matrix) with AllToAll operators, thus further reducing the communication volume. In Algorithm \ref{alg:optimized Tensor redistribution}, $InferSlice$ finds all necessary Slice-Op and appends them to the operator sequence $S$. If there is no more SliceOp,  $InferSlice$ sets $S\_flag$ to $False$. Similarly, $InferAll2All$ and $InferAllGather$ do the same things for AllToAllOp and AllGatherOp. 
Table \ref{tab: example of generatestrategyset} shows an example of using above mentioned three optimizations to fine-tune the redistribution sequence.

\begin{table*}[]
  \renewcommand{\arraystretch}{1.2} 
  \centering
  \caption{Tensor Redistribution Between $M_{from}=[-1, -1, 2, -1, 3]$ and $M_{to}=[1, -1, -1, 0,3]$} \label{tab: example of tensor redistribution}
  {
  \begin{tabular}{c|c|c|c}
  \hline
  \textbf{Step} & \textbf{Operation}          & \textbf{Tensor Map}    &\textbf{Communication Volume}          \\ \hline \hline
  \multirow{6}{*}{0: Initial}               & $AllGather(d_1, 1)$                 & $[-1, -1, 2, -1, 3]$ &\multirow{6}{*}{$\frac{d_1 d_2 d_3 - 1}{d_1 d_2 d_3}Size(T)$} \\ \cline{2-3} 
       & $AllGather(d_2, 2)$ & $[-1, -1, -1, -1, 3]$  \\ \cline{2-3} 
       & $AllGather(d_3, 4)$  & $[-1, -1, -1, -1, -1]$ \\ \cline{2-3} 
       & $Slice(d_1, 0)$     & $[1, -1, -1, -1, -1]$  \\ \cline{2-3} 
       & $Slice(d_0, 3)$     & $[1, -1, -1, 0, -1]$   \\ \cline{2-3} 
       & $Slice(d_3, 4)$     & $[1, -1, -1, 0, 3]$    \\ \hline \hline
  \multirow{4}{*}{1: Remove}             & $AllGather(d_1, 1)$                 & $[-1, -1, 2, -1, 3]$ &\multirow{4}{*}{$\frac{d_1 d_2 - 1}{d_1 d_2 d_3}Size(T)$}\\ \cline{2-3} 
       & $AllGather(d_2, 2)$ & $[-1, -1, -1, -1, 3]$  \\ \cline{2-3} 
       & $Slice(d_1, 0)    $ & $[1, -1, -1, -1, 3]$   \\ \cline{2-3} 
       & $Slice(d_0, 3)    $ & $[1, -1, -1, 0, 3]$    \\ \hline \hline
  \multirow{4}{*}{2: Rearrange}             & $Slice(d_0, 3)$                     & $[-1, 1, 2, 0, 3]$  &\multirow{4}{*}{$\frac{d_1+d_2 -2}{d_0 d_1 d_2 d_3}Size(T)$} \\ \cline{2-3} 
       & $AllGather(d_1, 1)$ & $[-1, -1, 2, 0, 3]$    \\ \cline{2-3} 
       & $Slice(d_1, 0)    $ & $[1, -1, 2, 0, 3]$     \\ \cline{2-3} 
       & $AllGather(d_2, 2)$ & $[1, -1, -1, 0, 3]$    \\ \hline \hline
  \multirow{3}{*}{3: Replace} & $Slice(d_0, 3)$                     & $[-1, 1, 2, 0, 3]$ &\multirow{3}{*}{$\frac{d_1 d_2 -1}{d_0 d_1^2 d_2 d_3 }Size(T)$}  \\ \cline{2-3} 
    & $AllToAll(d_1, 1 \rightarrow 0)$ & $[1, -1, 2, 0, 3]$   \\ \cline{2-3} 
    & $AllGather(d_2, 2)$ & $[1, -1, -1, 0, 3]$   \\ \hline
  \end{tabular}}
  \end{table*}

Finally, we obtain the inter-operator communication volume of such tensor redistribution by accumulating the communication volumes of redistribution operators within sequence $S$. 
Suppose we are using bandwidth optimal Ring-AllGather algorithm; the communication volume of AllGather is:
\begin{equation}
  V_{AG} = (device\_num - 1)data\_size;
\end{equation}
For AllToAll operators, each device only needs to send $\frac{1}{device\_num}$ different data to each other in the communication group. Thus, the communication volume of AllToAll is:
\begin{equation}
  V_{A2A} = \frac{(device\_num - 1)data\_size}{device\_num}.
\end{equation}
As we can see in Table \ref{tab: example of tensor redistribution}, the communication volume of the operator sequence that Algorithm \ref{alg:optimized Tensor redistribution} generates is much smaller. 

We then blend the bandwidth difference into the volume-based cost model to form our topology-aware cost model.

\begin{algorithm}
  \caption{Optimized Tensor Redistribution}
  \label{alg:optimized Tensor redistribution}
  \KwData{$\mathcal{D} = [d_{h-1}, d_{h-2}, ..., d_{0}]$; $M_{from} = [m_{from,0}, m_{from,1}, ..., m_{from, n-1}]$; $M_{to} = [m_{to,0}, m_{to,1}, m_{to, n-1}]$ 
  }
  \KwResult{Redistribution operator sequence $S$.}

  \While(){$M_{from} \neq M_{to}$}{
    $S\_flag \gets True$\;
    \While(){S\_flag}{
      $S\_flag \gets InferSlice(M_{from}, M_{to}, S)$\;

      $A2A\_flag \gets True$\;
      \While(){A2A\_flag}{
        $A2A\_flag \gets InferAll2All(M_{from}, M_{to}, S)$\;
      }
      }
    $AG\_flag \gets InferAllGather(M_{from}, M_{to},S)$\;
    \uIf{$AG\_flag == False$}{
      $AllGatherFirstUndoneDim(M_{from}, M_{to},S)$\;
    }
  }
\end{algorithm}

\subsection{Topology-aware cost model}
Based on volume-based cost model, we develop a topology-aware cost model that can additionally consider the bandwidth difference when calculating communicaiton costs. TAPS uses this topology-aware cost model to generate more fine-grained strategies.

\begin{figure*}
  \centering
  \includesvg[width=\linewidth]{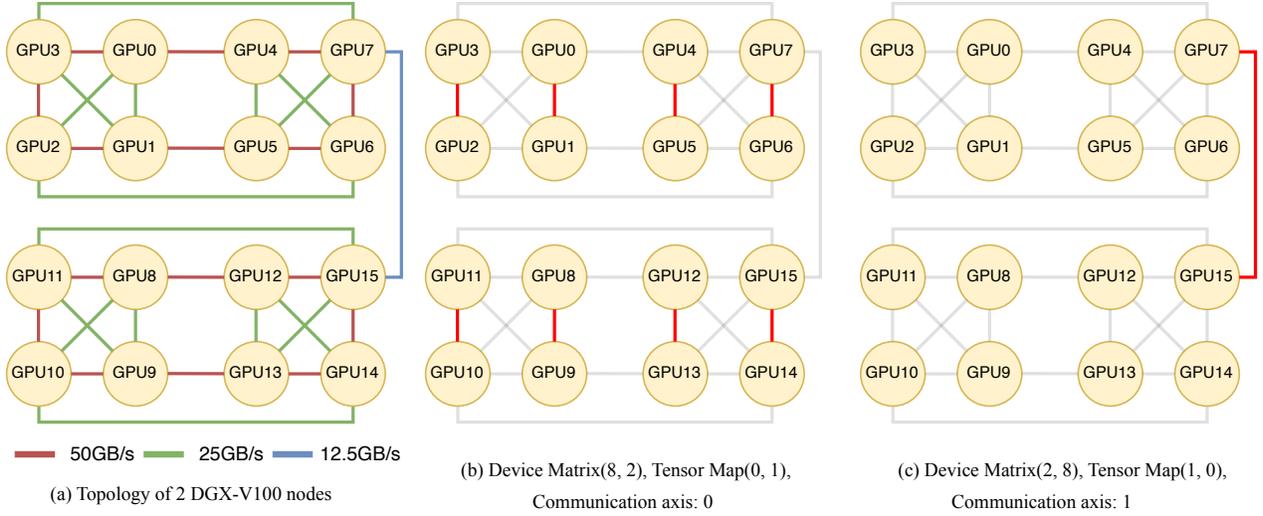}
  \caption{Intra-communication and Inter-communication on 2 DGX-V100 nodes}
  \label{fig: intra_comm vs inter_comm}
\end{figure*}

Our observation is that in multi-node environment, we can do multiple intra-node communications of different communication groups simultaneously, and they can all fully utilize the bandwidth; But for inter-node communications, they need to share the links between nodes, thus lowering the effective bandwidth of each communication group. Figure \ref{fig: intra_comm vs inter_comm}.(a) shows a 2 DGX-V100 nodes environment, where intra-node communication uses high-bandwidth NVLink and inter-node communication uses 100GBps InfiniBand. In Figure \ref{fig: intra_comm vs inter_comm}.(b), we do communication along the axis 0. There are 8 communication groups, which are (GPU0, GPU1), (GPU2, GPU3) and so on. Each of them has an individual NVLink to use and thus the effective bandwidth equals the bandwidth of NVLink. The case in Figure \ref{fig: intra_comm vs inter_comm}.(c) also has 8 communication groups, which are (GPU0, GPU8), (GPU1, GPU9) and so on. However, all of them need to transport data via the only inter-node link (i.e., red line in the figure). Since they are communicating simultaneously, we need to divide the bandwidth by 8. Thus, the effective bandwidth is $12.5/8 = 1.5625 GB/s$ in this case.

Based on this observation, TAPS computes the number the inter-node communication groups within a node for AllReduce, AllGather, and AllToAll operators to obtain the effective bandwidth for every communication group. TAPS computes the communication costs by dividing communication volumes by effective bandwidths.

\subsubsection{AllReduce}
\begin{algorithm}
  \caption{Infer number of inter-node communication groups within a  node for AllReduce}\label{alg:inter-communication}
  \KwData{Device Matrix $\mathcal{D}=({d_{h-1}, d_{h-2}, ..., d_0})$, Tensor Map $M=(m_0, m_1, ..., m_{s-1})$, device number in a node $local\_device\_num$}
  \KwResult{Number of inter-node communication group $ct$}
  $remain\_{devices} \leftarrow local\_device\_num$\;
  $Total\_devices \leftarrow product(\mathcal{D})$\;
  $parallel\_{degree} \leftarrow Total\_devices$\;
  $device\_{in} \gets 1$\;
  \lFor{$k \gets 0$ to $s-1$}
  { $parallel\_{degree} \leftarrow parallel\_{degree} \div d_{m_{k}}$}
  \For {$k \gets 0$ to $h-1$}{
    \uIf{ not $M.find(k)$ and $remain\_devices > 1$}
    {\uIf{$remain\_devices > d_{k}$}{
      $device\_{in} \gets device\_{in} \times d_{k}$\;
    }\uElse{
      $device\_{in} \gets remain\_devices$\;
    }        
    }  
    $remain\_devices \gets remain\_devices \div d_{k}$\;
  }
  \uIf{$device\_in \ge parallel\_degree$}{
    $ct \gets 0$\;}
  \uElseIf{$device\_in > 1$}{
    $ct \gets local\_device\_num \div device\_in $\;}
  \uElse{
    $ct \gets local\_device\_num$\;
  }
\end{algorithm}
For an arbitrary AllReduce operator, we first compute the inter-communication times of its input tensor using Algorithm \ref{alg:inter-communication}. Algorithm \ref{alg:inter-communication} takes the device matrix, tensor map of the communicated tensor, and the number of devices in a node as inputs, then infer the number of communication groups  that need to do inter-node communication. 

\begin{figure}
  \centering
  \includesvg[width=\linewidth]{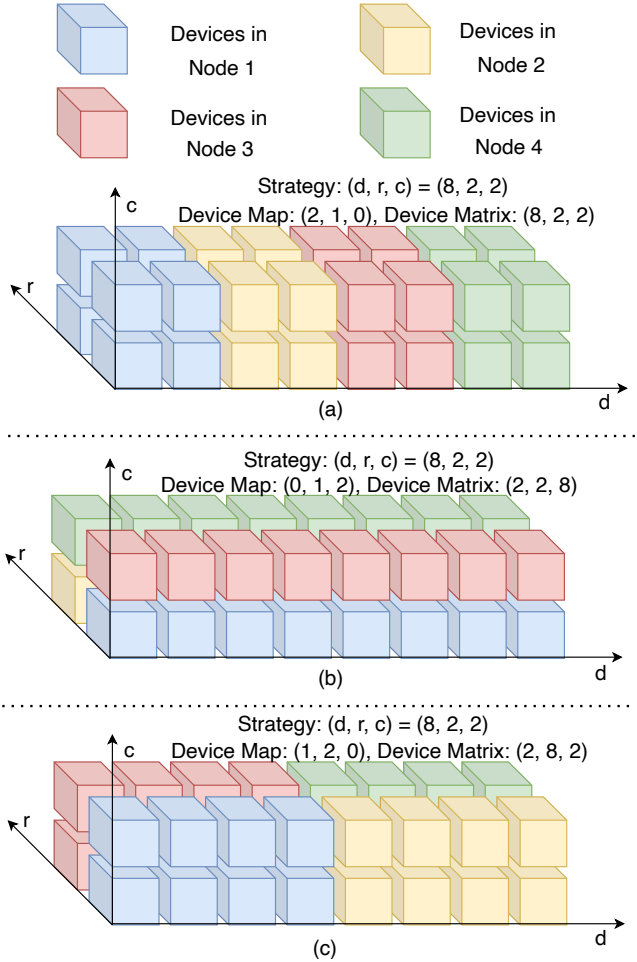}
  \caption{Device Distributions of Different Device Map and Device Matrix}
  \label{fig: device matrix}
 \end{figure}

The inter-communication times indicate how many communication groups do inter-node communications for a tensor simultaneously. For example, suppose we are executing a MatMul operator with strategy $(d, r, c)=(8,2,2)$ with different device maps as shown in Figure \ref{fig: device matrix}. The same-color cubes are devices within a node. In Figure \ref{fig: device matrix}(a), there are 4 different $\delta W$ partitions within a node, and they all need to communicate with other nodes. Thus, the inter-communication times $ct_{\delta W}$ is 4 in this case. The tensor $\delta W$'s inter-communication times $ct_{\delta W}$ are 4, 0, and 2 in Figure \ref{fig: device matrix}(a)(b)(c), respectively. 

Using the result of $ct$, we can compute the effective bandwidth $B_e$:
\begin{equation} \label{eq: effective bandwidth}
  \
  B_e = \left\{
    \begin{array}{lr}
      B_{intra} & ct_T = 0,  \vspace{2mm}\\
      B_{inter}/ct_T & ct_T > 0,
    \end{array}
  \right.
\end{equation}
where $B_{intra}$ is the intra-node communication bandwidth, and $B_{inter}$ is the inter-node communication bandwidth.

Finally, The communication cost of an AllReduce operator is:
\begin{equation}
  C_{AR} = V_{AR} / B_e.
\end{equation}

\subsubsection{AllGather}
Different from AllReduce, AllGather uses Algorithm \ref{alg:allgather inter-communication} to compute the inter-communication times of AllGather. Algorithm\ref{alg:allgather inter-communication} can compute the repetitive degree $r$ of AllGather in a device node and infer how many devices of a communication group are within a node. Using this information, it then outputs the $ct$ values of corresponding AllGather. TAPS also uses Eq.\ref{eq: effective bandwidth} to compute the $B_e$ for AllGather. The communication of an AllGather operator is:
\begin{equation}
  C_{AG} = V_{AG} / B_e.
\end{equation}

\begin{algorithm}
  \caption{Infer number of inter-node communication groups within a  node for AllGather}\label{alg:allgather inter-communication}
  \KwData{Device Matrix $\mathcal{D}=({d_{h-1}, d_{h-2}, ..., d_0})$, Tensor map $M=(m_0, m_1, ..., m_{s-1})$, Gather axis $g$, device number in a node $local\_device\_num$.}
  \KwResult{Inter communication times $ct$}
  $remain\_{devices} \leftarrow local\_device\_num$\;
  $parallel\_{degree} \gets d_{m_g}$\;
  $temp\_device\_num \gets 1$\;
  $repeat\_num \gets 1$\;
  \For {$k \gets 0$ to $g-1$}{
    $temp\_device\_num \gets temp\_device\_num \times d_{k}$\;
    \uIf{$not\ M.find(k) $}{
      $repeat\_num \gets repeat\_num \times d_{k}$\;
    }
    }
  \uIf{$repeat\_num > local\_device\_num$}{
    $repeat\_num \gets local\_device\_num$\;
  }
  \uIf{$temp\_device\_num \ge local\_device\_num $}{
    $ct \gets local\_device\_num \div repeat\_num$\;}
  \uElse{
    $remain\_devices \gets local\_device\_num \div temp\_device\_num$\;
    \uIf{$remain\_devices \ge parallel\_degree$}{
      $ct \gets 0$\;
    }
    \uElse{
      $ct \gets temp\_device\_num \div repeat\_num$\; 
    }
  }
   
\end{algorithm}

\subsubsection{AllToAll}
Unlike AllReduce and  AllGather, which utilize ring topology to communicate, AllToAll uses peer-to-peer (P2P) communication to exchange data within a communication group. While each node in AllReduce and AllGather has only one send link and receive link, each node in AllToAll establishes $p-1$ send and receive links that connect other nodes in the communication group, where $p$ is the device number of the AllToAll communication group. This may influence the communication volume we use to compute communication costs. Therefore, we need to recompute the communication volume for AllToAll. Suppose among $p$ devices, $k$ devices are within a node, and the tensor size is $T_S$. Then for any device in a node, there are $k-1$ intra-node communication with volume $T_S/p$, and $p-k$ inter-node communication with volume $T_S/p$. These $k$ devices accumulate established $(p-k)k$ connections to other nodes, each transport $T_s/p$ volume of data. Then for this AllToAll communication group, the inter-node communication volume via inter-node link is  $k(p-k) T_S/p$. Suppose there are $l$ devices in a node. Then there are $l/k$ different AllToAll communication groups within a node. We additionally suppose the repeat degree of them is $r$. The repetitive tensor slices could share the same communication results by synchronizing within a node using high-bandwidth links. Then in a node, $ct = l/(kr)$ groups simultaneously uses the inter-node bandwidth to communication, unless $p$ equals $k$. The values $k$ and $r$ can also be inferred using Algorithm \ref{alg:allgather inter-communication}. Effective bandwidth $B_e$ is computed using Eq.\ref{eq: effective bandwidth}. Thus, the communication cost of AllToAll is:

\begin{equation}
  \
  C_{A2A} = \left\{
    \begin{array}{lr}
      \frac{V_{A2A}}{B_e} & p = k,  \vspace{2mm}\\
      \frac{k(p-k)}{p-1}\frac{V_{A2A}}{B_e} & p > k,
    \end{array}
  \right.
\end{equation}

\section{Auxiliary Graph}
\begin{figure*}
  \centering
  \includesvg[width=\linewidth]{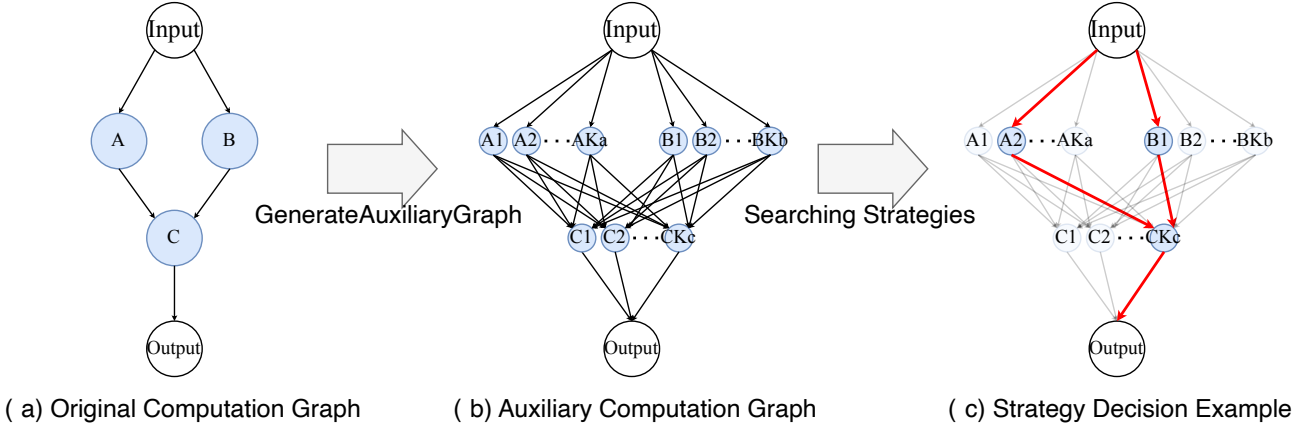}
  \caption{Auxiliary Computation Graph}
  \label{fig: auxiliary graph}
 \end{figure*}

 \begin{algorithm}
  \caption{GenerateAuxiliaryGraph}\label{alg:auxiliary_graph}
  \KwData{Computation graph $G=(V, E)$, device graph $D=(V_D, E_D)$}
  \KwResult{Auxiliary graph $G_A=(V_A, E_A)$}
  $V_A \leftarrow \emptyset, E_A \leftarrow \emptyset  $\;
  $device\_num \leftarrow |V_D|$\;
  \For{$ (u, w) \in E$}
  { $U_A$ = GenerateStrategySet($u$, $device\_num$)\;
    $W_A$ = GenerateStrategySet($w$, $device\_num$)\;
    $V_A = V_A \cup U_A \cup W_A$ \;
    \For {$u_a \in U_A$}
    {\For {$w_a \in W_A$}
    {
      $E_A = E_A \cup \{(u_a, w_a)\}$
    }
    }
  }
\end{algorithm}

Auxiliary graph $G_A=(V_A, E_A)$ is an extension of computation graph $G$ where each node $v_a \in V_A$ indicates a unique strategy of its original vertex $v \in V$. We use Algorithm \ref{alg:auxiliary_graph} to generate auxiliary graph as Figure \ref{fig: auxiliary graph}(a)(b) shows. For each $v_a \in V_A$, we label it with the original operator and a unique strategy.
The function GenerateStrategySet in algorithm \ref{alg:auxiliary_graph} enumerates all possible strategies of the input operator and creates corresponding auxiliary nodes for them. More specifically, GenerateStrategySet will generate $\sum_{i=1}^{\min{(p, n)}}i! \binom{p}{i} \binom{n-1}{i-1}$ different parallelism strategies when there is $p$ different partitionable axes in the operator and the operator is held by $N=2^n$ devices. A parallelism strategy of an operator consists of the parallelism degree and mapping of each axis. We can use the parallelism degrees to determine the partitions of each involved tensor of the operator and place them to corresponding devices according to the mappings.
For example, suppose a matrix multiplication (MatMul) operator that does computation $Y=XW$ can be partitioned along three axes: $b$ axis, $in$ axis, and $out$ axis; The unpartitioned shapes of $X$, $W$, and $Y$ are $(b, in)$, $(in, out)$, $(b, out)$, respectively. Table \ref{tab: example of generatestrategyset} shows the strategy set GenerateStrategySet generates when it takes the above MatMul operator and a device number of 4 as inputs. Taking $u_2$ in Table \ref{tab: example of generatestrategyset} for illustration, number 2 in $out$ axis represent tensor $W$ and $Y$ are sliced along $out$ axis into 2 parts; device map $(-1, 1, 0)$ indicates the mapping value of $b$, $in$, and $out$ axis are -1, 1, and 0, respectively. -1 here represents tensors replicated along the $b$ axis. 0 here indicates that the $out$ axis is partitioned most-innerly in clusters, which may have a high bandwidth when communicating. Device matrix $(1, 2, 2)$ is calculated by the parallelism degree of each axis and the device map, and it is a hierarchically logical topology of devices. 

After creating the vertices and edges of the auxiliary graph, we then compute the communication cost $C_{u_aw_a}$ of all $e_a=(u_a, w_a) \in E_A$ using our topology-aware cost model. The weight of $e_a$ equals the intra-operator cost of $w_a$ plus inter-operator cost between $u_a$ and $w_a$. 

Then, we can search strategies by selecting vertices and edges in the auxiliary graph. Figure \ref{fig: auxiliary graph} shows an example of the search result, where blue vertices are selected strategies.

\begin{table*}[]
\renewcommand{\arraystretch}{1.4}   
  \caption{Auxiliary nodes of a MatMul operator ($Y=XW$) partitioned on 4 devices generated by GenerateStrategySet}
    \label{tab: example of generatestrategyset}
    \centering
  \begin{tabular}{c|c|c|c|c|c|c|c|c}
  Node  & $b$ axis & $in$ axis & $out$ axis &X shape &W shape &Y shape &device map &device matrix  \\ \hline \hline
  $u_1$ & 1        & 1             & 4  & $(b, in)$ &$(in, out/4)$ &$(b,out/4)$     & (-1, -1, 0)   &(1, 1, 4)             \\
  $u_2$ & 1        & 2             & 2  & $(b, in/2)$ &$(in/2, out/2)$ &$(b,out/2)$ & (-1, 1, 0)    &(1, 2, 2)             \\
  $u_3$ & 1        & 2             & 2  & $(b, in/2)$ &$(in/2, out/2)$ &$(b,out/2)$ & (-1, 0, 1)    &(1, 2, 2)             \\
  $u_4$ & 1        & 4             & 1  & $(b, in/4)$ &$(in, out/4)$ &$(b,out)$     & (-1, 0, -1)   &(1, 4, 1)             \\
  $u_5$ & 2        & 1             & 2  & $(b/2, in)$ &$(in, out/2)$ &$(b/2,out/2)$ & (-1, 1, 0)    &(2, 1, 2)             \\
  $u_6$ & 2        & 1             & 2  & $(b/2, in)$ &$(in, out/2)$ &$(b/2,out/2)$ & (-1, 0, 1)    &(2, 1, 2)             \\
  $u_7$ & 2        & 2             & 1  & $(b/2, in/2)$ &$(in/2, out)$ &$(b/2,out)$ & (1, 0, -1)    &(2, 2, 1)             \\
  $u_8$ & 2        & 2             & 1  & $(b/2, in/2)$ &$(in/2, out)$ &$(b/2,out)$ & (0, 1, -1)    &(2, 2, 1)             \\
  $u_9$ & 4        & 1             & 1  & $(b/4, in)$ &$(in, out)$ &$(b/4,out)$     & (0, -1, -1)   &(4, 1, 1)            
  \end{tabular}
  \end{table*}

\section{Searching Strategies by ILP}
We formalize the strategy searching problem as an ILP problem as below shows:
\begin{alignat}{2}
  \min \quad &\sum_{(i, j) \in E_{A}}{B_{ij}C_{ij}} \label{eq: ILP} \\
  \mbox{s.t.} &\sum_{v_a \in V_{A}} X_{v_a}= 1, \quad \quad \quad \quad \quad \quad\quad\quad\quad\quad \forall v \in V \label{eq:one strategy constraint}\\
  &\sum_{(i, v_a)\in E_{A}}{B_{iv_a}} = X_{v_a}\times in\_degree(v) \nonumber \\
  &\sum_{(v_a,k)\in E_{A}}{B_{v_ak}} = X_{v_a}\times out\_degree(v), \forall v_a \in V_{A} \label{eq: in-out degree constraint}\\
  &\sum_{(i, j) \in E_{A}}{B_{ij}M_{ij} < Device\_Memory}, \label{eq:memory constraint}\\
  &B_{ij}, X_k \in \{0, 1\}, \quad \quad \quad \quad\forall (i, j) \in E_{A}, \forall k \in V_A\label{eq:binary}
\end{alignat}
where $X_{v_a}$, $B_{ij}$ are to-be-solved bool values that indicates the selection of the vertex $v_a\in V_A$ and edge $(i, j)\in E_A$. $C_{ij}$ and $M_{ij}$ are the communication and memory costs of edge $(i, j) \in E_A$. Equation \ref{eq:one strategy constraint} informs the solver that we only select one strategy for all $v\in V$. Equation \ref{eq: in-out degree constraint} limits any $v_a \in V_A$ to have the same indegree and outdegree as their original vertex $v \in V$. To avoid selecting multiple strategies for $v \in V$, we set the indegree and outdegree of $v_a \in V_A$ to zero if it is not selected. Equation \ref{eq:memory constraint} limits the solver to produce overall strategies that do not exceed device memory. 

Instead of dynamic programming, we use integer linear programming for two reasons. First, dynamic programming methods like \cite{optcnn, wang2019tofu} cannot capture the overall memory cost during processing, which might generate strategies that exceed memory constraints. Although methods like \cite{cai2021tensoropt} maintain a communication-memory-cost bound to avoid this drawback, its computation complexity is unacceptable while generating strategies for large-scale models. Second, we can directly use a high-performance third-party solver to solve the ILP problem, which saves our time from optimizing the solver runtime. 

\section{Evaluation}
We evaluate TAPS by comparing the communication costs of strategies generated by volume-based searching and topology-aware searching. In our evaluation, we assume the intra-bandwidth equals 60GB/s and the inter-bandwidth equals 6GB/s. These two results are the peak bandwidth we get after testing on two 8-V100 nodes using nccl-tests\cite{nccltest}. Additionally, we assume that all the communication can fully utilize the bandwidth and that we are running in a homogeneous environment.

\subsection{Searching Runtime}
We test the searching runtime on searching strategies for AlexNet\cite{alexnet} and Megatron-LM\cite{megatron2021, megatron2019}. Note that the main body of transformer-based networks consists of several layers with the same structures. Given that the same structures always have the same strategies when the devices they use are homogeneous, we only search strategies for one transformer layer of the networks. The applied solver can solve strategies within a few seconds using a 16-core 3.2GHz Intel i9-12900K CPU. In our searching runtime experiment, we suppose each node has 8 devices.

Table \ref{tab: search runtime} shows some examples of running time of solving strategies, where $V_D$ is the total number of devices, and $|E_A|$ is the number of auxiliary edges. The search time is irrelevant to the number of a model's parameters. Instead, it is relevant to the number of a model's operators and the total device number. For example, the transformer layer of Megatron-LM 1.7B and 3.6B has the same structure but different parameter numbers. We follow the configurations in \cite{megatron2021}, searching intra-operator strategies for 1.7B, 3.6B both on overall 32 devices. As Table \ref{tab: search runtime} shows, the $|E_A|$ and their time remain in the same order of magnitude.

\begin{table}[]
\centering
\caption{Strategy Searching Time of Solver} \label{tab: search runtime}
\resizebox{\linewidth}{!}{
\begin{tabular}{c|c|c|c}
\textbf{Model}   & \textbf{$|V_D|$} & \textbf{$|E_A|$} & \textbf{Time} \\ \hline \hline
AlexNet          & 8                        & $>3\times 10^3$                     & $< 0.1$s      \\ 
AlexNet          & 16                        & $>10^4$                     & $< 0.2$s      \\ 
AlexNet          & 64                        & $>5\times 10^4$             & $<0.8$s       \\ 
Megatron-LM 1.7B & 32                        & $>3\times 10^4$             & $<0.4$s       \\ 
Megatron-LM 3.6B & 32                        & $>3\times 10^4$             & $<0.4$s       \\ 
Megatron-LM 1T   & 8                         & $>4\times 10^3$             & $<0.1$s       \\ 
\end{tabular}}
\end{table}

\begin{figure*}
    \centering
    \includesvg[]{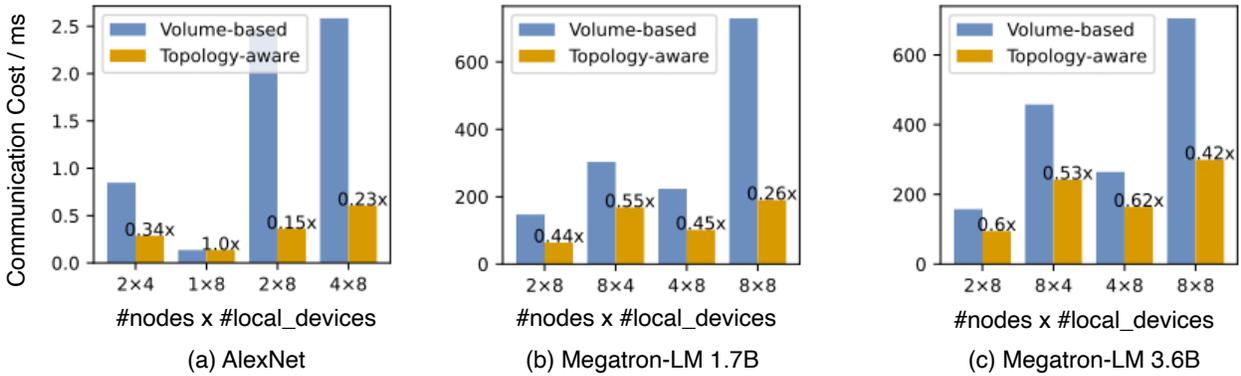}
    \caption{Comparison of Topology-Aware and Volume-Based Searching on Different Models}
    \label{fig: comparison}
\end{figure*}

\begin{figure}
  \centering
  \includesvg[width=\linewidth]{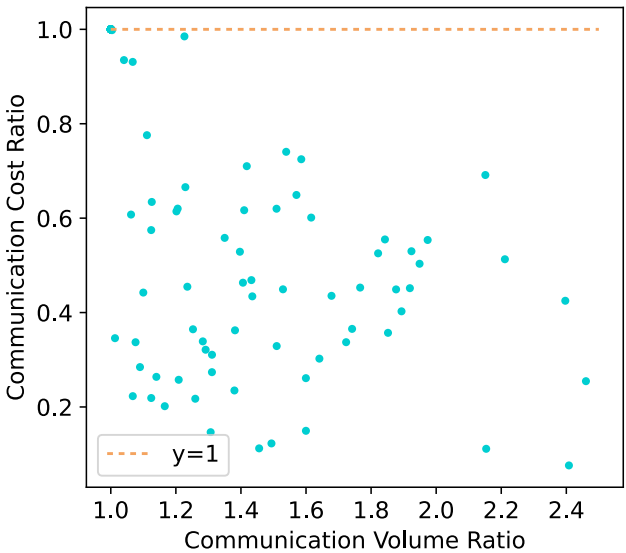}
  \caption{Ratios of the Topology-Aware and Volume-Based Searching}
  \label{fig: ratio}
\end{figure}

\subsection{Comparison with Volume-Based Searching}
We compare the communication cost between strategies that volume-based searching and topology-aware searching solve. In our experiment, we search strategies for the convolution network AlexNet, and transformer-based networks Megatron-LM. We do the volume-based searching by replacing the communication costs of the auxiliary edges with their corresponding communication volumes, after which we use the same solver to search for the strategies. Then we compute the communication costs of generated volume-based searching results using our topology-aware cost model.
The comparison results are shown in Figure \ref{fig: comparison}, where blue bars are the communication costs of volume-based searching results, and orange bars are the communication costs of topology-aware searching results. We take AlexNet, Megatron-LM 1.7B, and 3.6B as examples. As we can see in \ref{fig: comparison}(a), when there is only one node, the communication cost of two different search results will be the same. This is intuitive since no inter-node communication exists in this case. Our experiments show that TAPS can always find strategies that outperform those volume-based searching solve out. In the case of searching strategies for AlexNet on two 8-device nodes, it even reduces the communication cost by 85\%.
Additionally, we merge all experiments we run into Figure \ref{fig: ratio}, where each point represents a search of a model under a specific device topology. The $x$-axis represents the ratio of topology-aware communication volume and volume-based communication cost; The $y$-axis represents the ratio of topology-aware communication cost and volume-based communication cost. As we can see, all points in the graph lie on or below the line $y=1$, which means that topology-aware searching can always find strategies with smaller communication costs than volume-based searching. Moreover, topology-aware searching reduces the communication cost by more than 20\% in most cases.

\section{Related Work and Discussion}
\textbf{Pipeline Parallelism.} auto parallleism methods like Alpa, Chimera\cite{Chimera} and PipeDream\cite{PipeDream} can generate pipeline parallelism strategies that balance the stages on different devices. The searching space of TAPS is orthogonal to pipeline parallelism. Thus we can use TAPS to search intra-operator parallelism strategies for each stage of the pipeline.

\textbf{Multi-dimensional Tensor Parallelism.} 2D-TP\cite{optimus}, 3D-TP\cite{Youyang-3D} from Colossal-AI\cite{Colossal-AI} generate intra-operator strategy heuristically. TAPS currently does not support  2D-TP because 2D-TP uses Broadcast and Reduce to finish the communication, while we use AllGather, AllToAll, and Slice instead. TAPS naturally includes strategies of 3D-TP because 

\textbf{Overlapping Communication and Computation.}
In our implementation, we assume that the communication cannot overlap with computation; thus, we can ignore the computation costs. However, in actual training cases, researchers\cite{jangda2021breaking} delegate to overlap the computation and communication. It is hard for us to be aware of the overlap degree. A trade-off solution is manually setting the overlap degree for communications of different dimensions. For example, in some cases, the communication of data parallelism can be fully overlapped, then we can set the overlap degree to 1.

\textbf{Estimating the Costs using regression models.}
Although we assume the bandwidth can be fully utilized, we notice that the effective bandwidth is very low when the size of transferred data is small. This is because, during communication, there are overheads like creating connections and computing average values. Using regression models to simulate the variations of effective bandwidth is a good choice to improve TAPS further.

\section{Conclusion}
We present TAPS, a topology-aware intra-operator parallelism strategy searching algorithm that generates fine-grained intra-operator strategies for multi-node environments. TAPS can generate tensor redistribution operations with fewer communication costs heuristically. TAPS calculates the communication costs of each strategy according to communication volume and effective bandwidth, thus producing more reasonable strategies compared to methods that only consider communication volume. Based on the communication costs, TAPS formalizes the searching problem as an integer linear programming problem by creating and utilizing an auxiliary graph and then solving the result within a few seconds. Compared to volume-based searching algorithms, TAPS can generate strategies with up to 85\% fewer communication costs for cases in multi-node environment. The source code of TAPS will be publicly available.

\bibliographystyle{plain}
\bibliography{references}

\end{document}